\documentclass{aa} 
\usepackage{natbib}
\usepackage[dvips, pdftex]{graphicx}
\usepackage{epstopdf}
\usepackage{txfonts}
\bibpunct{(}{)}{;}{a}{}{,}

\begin{document}
\authorrunning{Hu et al.}

   \title{A seismic approach to testing different formation channels of subdwarf B stars}

   \author{Haili Hu
          \inst{1,2}
          \and
          M.-A.~Dupret\inst{3}
          \and               
          C.~Aerts\inst{1,2}
          \and
          G.~Nelemans\inst{1}
          \and                                            
          S.~D.~Kawaler\inst{4}
          \and  
          A.~Miglio\inst{5}
          \and
          J.~Montalban\inst{5}
          \and 
          R.~Scuflaire\inst{5}
         }

%   \offprints{H.~Hu}

   \institute{Department of Astrophysics, IMAPP, Radboud University Nijmegen, PO Box 9010, 6500 GL, Nijmegen, the Netherlands\\
              \email{hailihu@astro.ru.nl}
         \and
             Institute of Astronomy, Katholieke Universiteit Leuven, Celestijnenlaan 200D, 3001 Leuven, Belgium
          \and
          Observatoire de Paris, LESIA, 5 place Jules Janssen, 92195 Meudon Principal Cedex, France
          \and{Department of Physics and Astronomy, Iowa State Univeristy, Ames, IA 50014, USA }
          \and
          Institute d'Astrophysique et G\'eophysique, universit\'e de Li\`ege, Belgium\\
             }

   \date{Received 21 May 2008 / Accepted 22 July 2008 }

% \abstract{}{}{}{}{} 
% 5 {} token are mandatory
 
  \abstract
  % context heading (optional)
  % {} leave it empty if necessary  
   {There are many unknowns in the formation of subdwarf B stars. Different
   formation channels are considered to be possible and to lead to a variety of
   helium-burning subdwarfs. All seismic models to date, however, assume that
   a subdwarf B star is a post-helium-flash-core surrounded by a thin inert layer
   of hydrogen.}
  % aims heading (mandatory)
   {We examine an alternative formation channel, in which the subdwarf B star
   originates from a massive ($>$ $\sim$2 M$_{\odot}$) red giant with a
   non-degenerate helium-core. Although these subdwarfs may evolve through the same
   region of the $\log g-T_{\rm eff}$ diagram as the canonical post-flash
   subdwarfs, their interior structure is rather different. We examine how this
   difference affects their pulsation modes and whether it can be observed.}
  % methods heading (mandatory)
   {Using detailed stellar evolution calculations we construct subdwarf B models
   from both formation channels. The iron accumulation in the driving region due
   to diffusion, which causes the excitation of the modes, is approximated by a
   Gaussian function. The pulsation modes and frequencies are calculated with a
   non-adiabatic pulsation code. }
  % results heading (mandatory)
   {A detailed comparison of two subdwarf B models from different channels, but
with the same $\log g$ and $T_{\rm eff}$, shows that their mode excitation is
different. The excited frequencies are lower for the post-flash than
for the post-non-degenerate subdwarf B star. This is mainly due to the differing chemical
composition of the stellar envelope. A more general comparison between two grids
of models shows that the excited frequencies of most post-non-degenerate
subdwarfs cannot be well-matched with the frequencies of post-flash subdwarfs.
In the rare event that an acceptable seismic match is found, additional
information, such as mode identification and $\log g$ and $T_{\rm eff}$
determinations, allows us to distinguish between the two formation channels. }
% conclusions heading (optional), leave it empty if necessary 
{}


   \maketitle

\section{Introduction}
Commonly, subdwarf B (sdB) stars are identified as extreme horizontal branch
(EHB) stars, and they are believed to be post-He-core-flash products with core
masses $\sim$0.5 M$_{\odot}$ surrounded by a very thin inert H-envelope \citep{heber1986,saffer1994}. From a single stellar evolution
point of view, this can be explained by enhanced mass loss of stars close to
He-ignition with very lightly bound envelopes \citep{d'cruz1996}, i.e. stars
with degenerate cores near the tip of the red giant branch (RGB). However, as
they are frequently observed in binaries
(e.g.~\citealt{allard1994,morales2006}), binary interactions most likely play an
important role in their formation. \citet{han2002} explored the main binary
evolution channels that can produce sdB stars: common-envelope ejection (CEE),
stable Roche lobe overflow (RLOF), and helium white dwarf mergers. They
found that the sdB mass distribution may be much broader than previously
thought, $0.3-0.8$ M$_{\odot}$ instead of $0.4-0.5$ M$_{\odot}$. The sdB stars
with non-canonical masses follow from mergers or massive ($>$ $\sim$2
M$_{\odot}$) progenitors that ignite helium quiescently, where the latter can be
a subchannel of either the CEE channel or the stable RLOF channel. Binary
population synthesis shows that the massive progenitors do
not contribute significantly to the sdB population \citep{han2003}. But one
should keep in mind that it is asumed in such studies that CE evolution is
described by the $\alpha$ formalism, i.e. that the CE ejection is driven by the
orbital energy. Because the physics of the CE phase is poorly understood, other
scenarios should not be excluded a priori. For example, the $\gamma$-formalism
proposed by \citet{nelemans2000}, based on the angular momentum equation rather
than the energy equation, provides an alternative description. In this case, the
massive red giants cannot be ruled out as possible progenitors of post-CE sdB stars
\citep{hu2007}. We therefore want to explore the possibility of this neglected
class of progenitors in a different manner, by using the seismic properties that
have been observed in some sdB stars.

Although the post-flash and the post-non-degenerate sdB stars can appear in
the same $\log g-T_{\rm{eff}}$ region, their interior structure is quite
different. In particular, the chemical composition profiles differ greatly
depending on whether helium ignited in a flash or quiescently. For example, the
canonical post-He-flash sdB star has a very narrow He$-$H transition zone, while
the sdB star created from a more massive progentior has a much broader
H-profile. This is a direct result of the differing chemical compositions
between low-mass and high-mass stars on the RGB, owing to the
different convective regions during the main-sequence and RGB evolution. We examine whether this difference in the interior structure will result in
observable differences in the pulsation modes.

The sdB pulsators consist of two classes, the short-period variable EC 14026
stars \citep{kilkenny1997}, and the long-period variable PG 1716 stars
\citep{green2003}. The rapid oscillations in EC 14026 stars are interpreted in
terms of low-order $p$-modes \citep{charpinet1996}, driven by the
$\kappa$-mechanism operating in the iron opacity bump. The same mechanism
has been shown to excite long-period, high-order $g$-modes in the cooler models
\citep{fontaine2003}. The local iron enhancement necessary in the driving region
around $\log T\approx 5.3$ is due to the competing diffusion processes of
radiative
levitation and gravitational settling. It is well-known that the opacities play
an important role in the study of the pulsations. \citet{seaton2004} showed that
the iron opacity bump is situated at slightly higher temperatures using OP opacities
\citep{seaton1994,badnell2003} compared with OPAL opacities
\citep{iglesias1996}. \citet{jeffery2006} found that, using OP opacities and
nickel enhancement in addition to iron, the theoretical instability
strip of $g$-mode sdB oscillators is more consistent with observations. For our
purposes it is sufficient to use OPAL opacities and iron enhancement,
since we are interested in the relative differences between two types of sdB
stars. We acknowledge the importance of including the effect of OP opacities and
nickel enhancement in further detailed studies.

The details of the computations are given in \S \ref{computations}. The
results are presented in \S \ref{results}. In \S \ref{structure}, we
compare the detailed physical characteristics of two reference models with
different formation histories. In \S \ref{grid} we compare the frequency
characteristics globally between two grids of models. The results and conclusions
are discussed in \S \ref{discussion}.

\section{Computations}\label{computations}
\subsection{The evolution calculations}\label{ev}
We constructed sdB structure models with the stellar evolution code developed by
\citet{eggleton1971, eggleton1972, eggleton1973}, \citet{ faulkner1973}, and updated by
\citet{han1994} and \citet{pols1995,pols1998}.
The updated version of the code uses an equation of state
that includes pressure ionization and Coulomb interaction, nuclear reaction
rates from \citet{caughlan1985} and \citet{caughlan1988}, and neutrino loss
rates from \citet{itoh1989,itoh1992}. Both convective and semi-convective mixing
are treated as diffusion processes. It is assumed that mixing occurs in regions where
\begin{equation}
\nabla_{\rm rad}>\nabla_{\rm ad}-\delta_{\rm ov}/(2.5+20\beta+16\beta^2),
\end{equation} 
where $\beta$ is the ratio of radiation pressure to gas pressure and
$\delta_{\rm{ov}}$ is the overshooting parameter. \citet{schroder1997} showed
that $\delta_{\rm{ov}}=0.12$ gives the best fit to observations of $\zeta$
Aurigae binaries, which corresponds to an overshooting length of $\sim$$0.25
H_p$. For our comparative study, it suffices to adopt $\delta_{\rm{ov}}=0.12$,
but keep in mind that core overshooting can in fact also be probed by
asteroseismology, e.g.~\citet{aerts2003}.

We evolved stars assuming a chemical composition of $X=0.70$ and $Z=0.02$. We used  a mixing-length parameter (the ratio of the mixing-length to the local pressure scaleheight) of $\alpha=l/H_p=2.0$. If not mentioned
otherwise, the opacity tables were constructed by combining the OPAL opacities
\citep{iglesias1996} with the conductive opacities \citep{hubbard1969,
canuto1970,iben1975}, as implemented in the Eggleton code by
\citet{eldridge2004}.

We started by evolving zero-age main-sequence (ZAMS) models in the range $1-3$ M$_{\odot}$ to the tip of
the RGB, adopting a Reimer's mass loss rate \citep{reimers1975},
\begin{equation}
\dot{M}_{\rm{wind}}=4\times10^{-13}\eta \frac{(R/R_{\odot})(L/L_{\odot})}{(M/M_{\odot})}\textrm{  [M}_{\odot}\textrm{yr}^{-1}],
\end{equation}
with an efficiency of $\eta=0.4$ \citep{iben1983,carraro1996}.  For simplicity, we did not include mass-loss on the EHB. \citet{unglaub2001} showed that, if the
observed chemical abundances are the result of the combined effects of diffusion
and mass loss, the sdB mass-loss rate should be in the range $10^{-14}\leq
\dot{M} ($M$_{\odot}$yr$^{-1})\leq 10^{-12}$. This is consistent with the rates
found by \citet{vink2002} for radiation-driven wind models. They also showed
that these rates are too low to have a direct effect on the sdB evolution.

At the RGB tip, we removed the envelope, while keeping the chemical compositions
fixed. Thus, we assume that the mass transfer happens on a much shorter timescale than the nuclear timescale. This is a reasonable assumption for sdB
stars in short-period binaries formed by CE ejection, which is the majority of
the observed sdB stars \citep{maxted2001} and the focus of our study here. In
the case that the He-flash occurs, zero-age horizontal branch models were
artificialy created from a 2.25 M$_{\odot}$ He-core-burning star, where we reset the chemical compositions to the values before the flash. This treatment is not rigorously valid. Full evolutionary models of the He-flash show that the C abundance in the He-core can increase up to $\sim$5\% \citep{piersanti2004, serenelli2005}. Since the $p$-modes are not sensitive to the core, we are not worried about this. An interesting scenario is an sdB star that is formed by a late He-core flash on the white dwarf cooling curve \citep{castellani1993}. In such a case the He-flash-driven convection zone can penetrate into the H-rich layers, resulting in a surface enrichment of He and C \citep{ brown2001, schlattl2001, cassisi2003}. We note that this might influence the pulsations, but we will not discuss this scenario further here.

On the EHB, we used for temperatures $\log T>7$ the same opacities as
mentioned above.  In the outer layers of the star, $\log T<7$, where the
pulsation driving region is located, the opacities were calculated by
interpolating between several OPAL tables computed with iron abundance enhanced
by factors of $f=1$, 2, 5 , and 10 relative to solar, thus $X(Fe) =
0.071794Zf$. The abundances of the other heavy elements are decreased such that
the overall metallicity is kept constant as in \citet{miglio2007}.

\subsection{The oscillation calculations and iron accumulation}\label{osc}
We adapted the Eggleton evolution code so that the output is suitable for
pulsation calculations. In practice, this implied calculating some additional
physical quantities during the evolution, and modifying the mesh to have
sufficient meshpoints in the stellar envelope. The seismic properties of the
stellar models are then calculated with two pulsation codes. The adiabatic code
OSC by \citet{scuflaire2007} is used to obtain the approximate frequencies, which
are used as a first guess in the linear non-adiabatic code MAD by
\citet{dupret2001}. We determined the theoretical frequency spectrum up to $l=2$,
since it is expected that higher order modes are geometrically cancelled.
\citet{charpinet1996} has established that the excitation of sdB oscillatons is
related to a local enrichment of iron in the stellar envelope caused by
diffusion. Radiative levitation is expected to set up significant chemical
gradients within a diffusion timescale of $\sim$$10^5$ yr, and consequently iron
accumulates around $\log T\approx 5.3$ \citep{michaud1989,
chayer1995}. Time-dependent diffusion calculations \citep{fontaine2006} show
that, after $\sim$$10^5$ yr, many pulsation modes are excited. Since element
diffusion is not treated in the evolution code, we used an approximation for the
iron accumulation, assuming that the iron only affects the stellar
structure through the opacity. At each timestep of the evolution calculations,
the iron enhancement factor $f$ is increased with a Gaussian centered at $\log
T=5.3$,

\begin{equation}\label{df/dt}
\frac{d f}{dt} = \frac{(1- f /10)^3}{\tau}\exp\big(-\frac{(\log T-5.3)^2}{\sigma^2}\big),
\end{equation}
with the initial condition $f(t=0)=1$. The width $\sigma^2 = 0.05$ and
accumulation timescale $\tau = 4\times 10^5$ yr are chosen such that iron is
only increased in the region $4.5<\log T<6.1$, and $\lim_{t\rightarrow \infty}
f=10$, which is loosely based on the time-dependent diffusion calculations of
\citet{fontaine2006} and the equilibrium profiles of \citet{charpinet1997}. Our
parametric approximation is rather ad hoc, but since we are interested in the
relative differences between two different scenarios, the exact shape of the
iron profile is not crucial here. We will discuss the influence of the iron profile 
on the pulsations and the evolution of the star in \S \ref{ironacc}.

\section{Results}\label{results}
\subsection{Effects of the iron accumulation}\label{ironacc}
\begin{figure}
   \begin{center}
  \includegraphics[angle=-90, width=9cm]{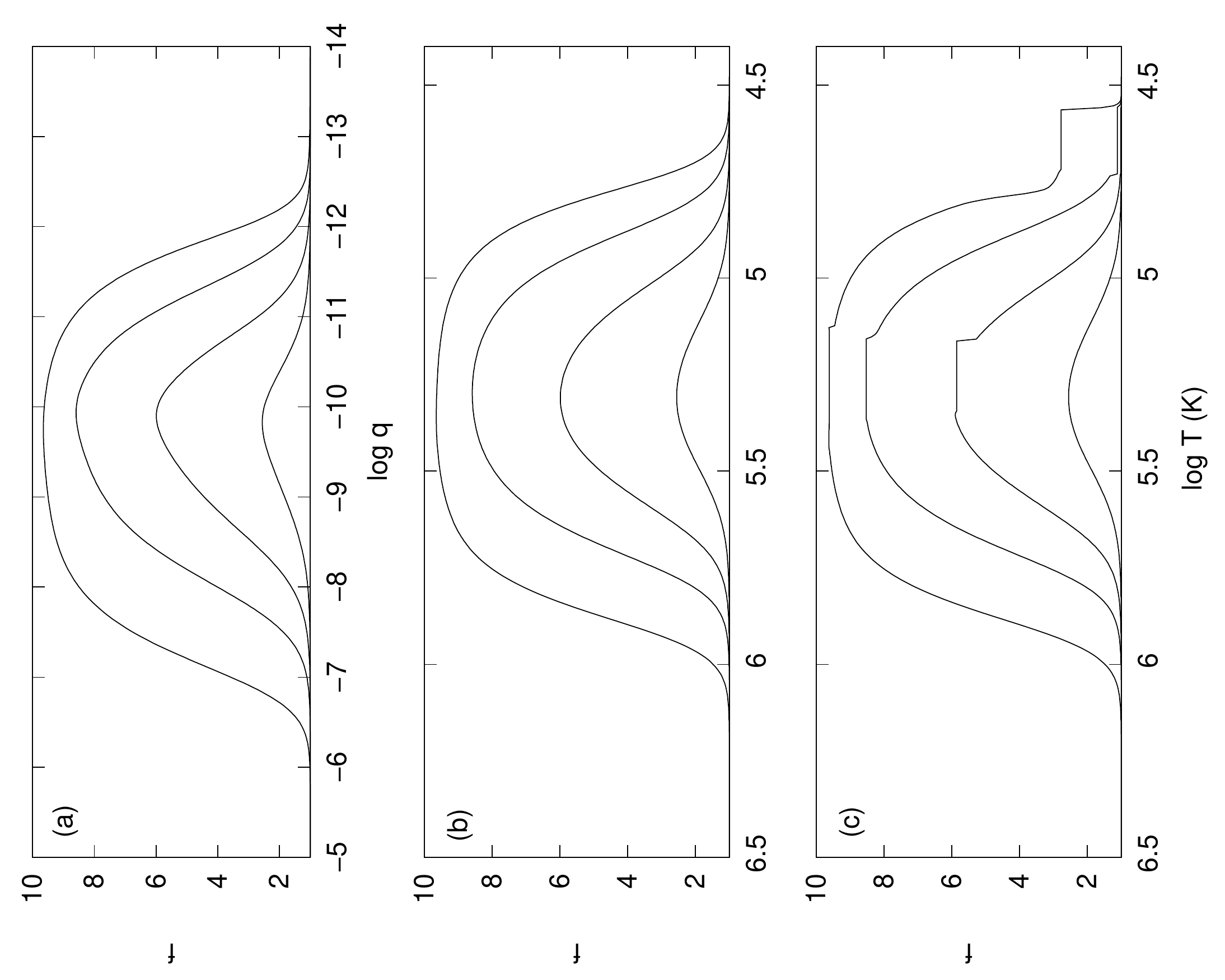}
   \caption{The iron enhancement factor $f$ throughout the star as a function of
   (a) the outer mass fraction: $\log q=\log\big(1-\frac{M_r}{M_*}\big)$, and
   (b) the temperature. In panel (c), we show the effect of convective mixing on
   our parametric iron profile. The sdB star has $M_{\rm core}=0.47$ M$_{\odot}$
   and $M_{\rm env, 0}= 10^{-4}$ M$_{\odot}$, and was constructed from
   a $1.00$ M$_{\odot}$ ZAMS model. The profiles from bottom to top correspond
   to sdB ages $10^5$, $10^6$, $10^7$, $1.8\times10^8$ yr, where the last model
   is at the end of core-He-burning. }
              \label{fprofiles}
              \end{center}
    \end{figure}

In Fig.~\ref{fprofiles}, we show $f$ throughout the star for different ages of
an sdB star. Note that the temperature range $4.5<\log T<6.1$ corresponds to a
very narrow mass shell of $\sim$$10^{-6}$ M$_{\odot}$. In Fig.~\ref{fprofiles}c,
we included the backreaction of convective mixing on our
parametric iron profile during the evolution. Note   that the iron abundance is homogeneous near
$\log T=5.3$ ($\log q = -10$) and $\log T=4.6$ ($\log q = -12.5$). This is
caused by two narrow convective layers due to iron and helium ionization,
respectively. Interestingly, we found that the convective region around $\log
T=5.3$ would not be present without iron accumulation. We determined that the
slightly perturbed iron profile has a negligible effect on the driving and the
pulsation frequencies. Moreover, since our description of iron accumulation is approximate, we have not included this effect in subsequent calculations.

%______________________________________________ 
%______________________________________________ 

\begin{figure}
   \begin{center}
  \includegraphics[angle=-90, width=9cm]{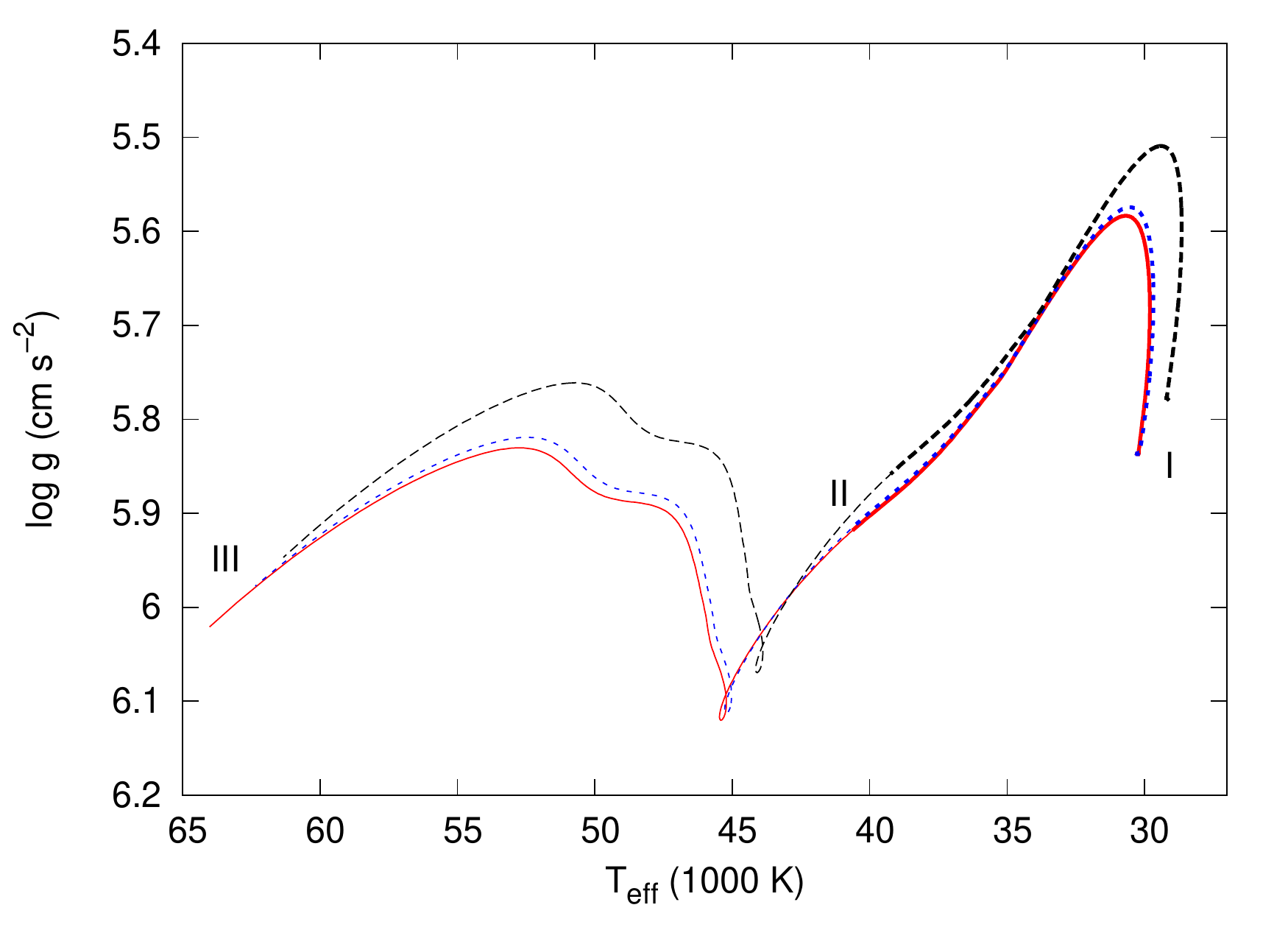}
   \caption{Evolutionary tracks in the $\log g-T_{\rm eff}$ diagram of sdB stars
   with different iron profiles. The three tracks start with the same zero-age EHB model
   with $M_{\rm sdB}=0.47$ M$_{\odot}$ and $M_{\rm env, 0}=10^{-4}$ M$_{\odot}$,
   created from a $1.00$ M$_{\odot}$ ZAMS model. The lowest solid curve is for a
   model with no iron enhancement, the middle dotted curve is for a Gaussian
   iron increase centered at $\log T=5.3$, and the upper dashed curve is for iron
   increased with a factor 10 uniformly in the envelope. The labels I, II, II
   indicate the ZAHB, the end of core-He burning and the end of He-shell burning,
   respectively. It takes the sdB star $1.8\times 10^8$ yr to evolve from I to
   II, and $10^7$ yr from II to III. }
              \label{fcompare}
              \end{center}
    \end{figure}

In Fig.~\ref{fcompare}, we show the effect of different iron abundance profiles
on the sdB evolution in the $\log g-T_{\rm eff}$ diagram. We compared our
parametric approach (Eq.~\ref{df/dt}) with the case of no iron enhancement, and a uniform
enhancement of $f=10$ in the whole envelope, as used in studies of mode
excitation \citep{jeffery2006}. It is evident that using $f=10$ influences the
evolution drastically, while our Gaussian parametrization of $f$ has little effect. Higher
iron abundances indeed give higher opacities, and thus larger stellar
radii. With Eq.~\ref{df/dt} we only increase iron in a relatively
small region, resulting in a minimal effect on the stellar structure. However, a
minimal change of the stellar structure can result in a visible shift of the
pulsation frequencies, as shown by \citet{fontaine2006}.

\begin{figure}
   \begin{center}
  \includegraphics[angle=-90, width=9cm]{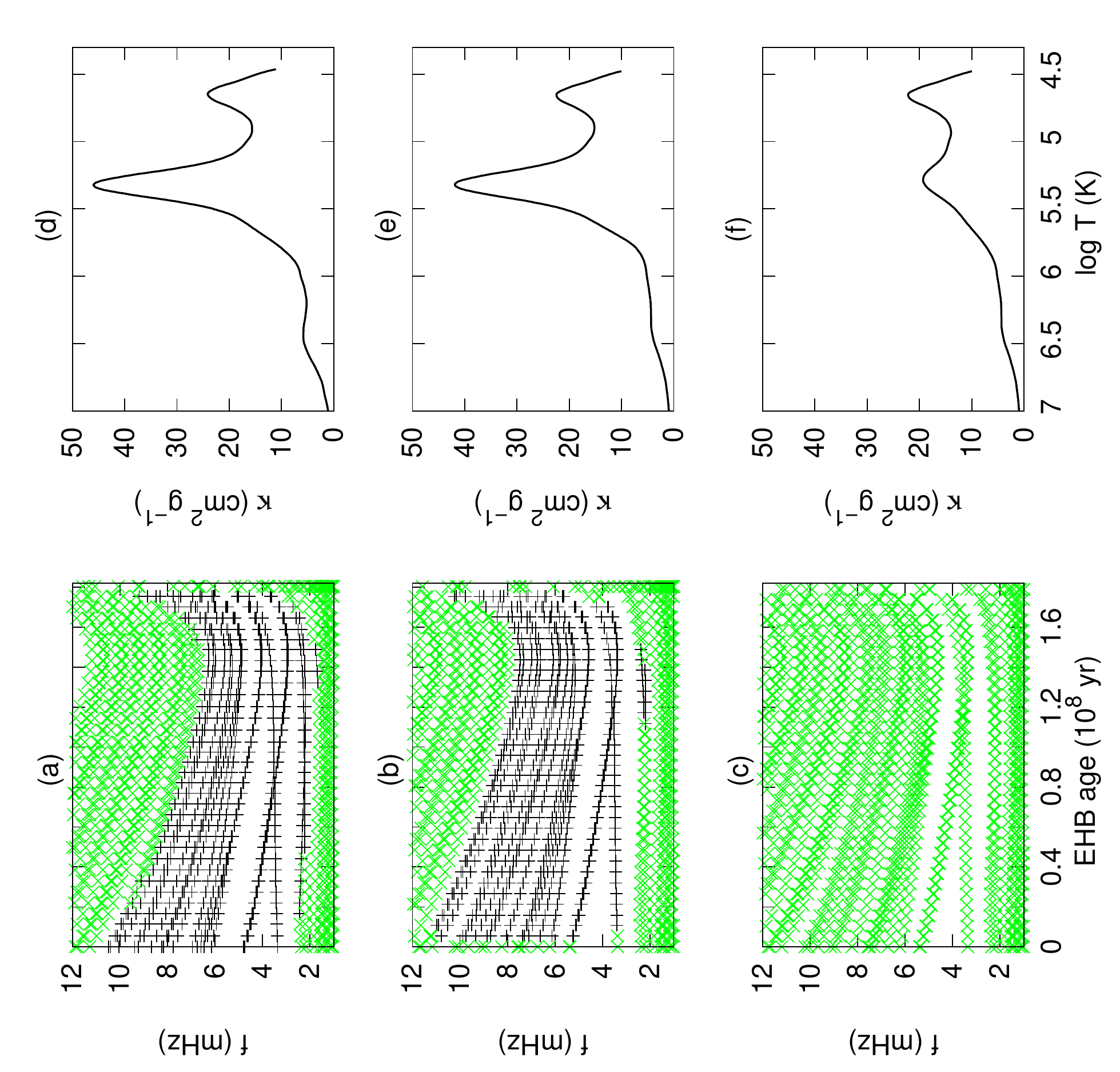}
   \caption{Left: the frequencies of the stable ($\times$) and unstable ($+$)
   modes ($l\leq 2$) as a function of the EHB age, for the three tracks
   given in Fig.~\ref{fcompare}. Right: the opacity as a function of the
   temperature after $10^7$ yr of sdB evolution. Panels (a\&d) correspond to
   $f=10$, (b\&e) are for the Gaussian parametrization of $f$, and (c\&f) are for the model
   with no iron enhancement.}
              \label{mode_kappa}
              \end{center}
    \end{figure}

We also compared the effect on the excitation, and find that our approximation of the 
iron profile can excite almost as many modes as the $f=10$ enhancement, see
Fig.~\ref{mode_kappa}a-c. This is understood in terms of 
the driving mechanism in sdB stars, which is
associated with the iron opacity bump such that 
accumulating iron in this driving
region is sufficient for the excitation of the pulsation modes, see
Fig.~\ref{mode_kappa}d-f. Thus, with our parametric approach to iron
accumulation, the issue of excitation can be addressed while keeping the effect
of the iron profile on the stellar structure realistic.
Furthermore, these are the first evolutionary models of sdB stars that
include the effect of iron accumulation, albeit in an approximative manner.

%______________________________________________ 
%______________________________________________ 
%______________________________________________ 

\subsection{The stellar models}
Following the procedure as described in \S \ref{computations}, we constructed a grid of canonical,
i.e.~post-He-flash, sdB models (hereafter called grid $A$) with masses in the
range $0.42-0.47$ M$_{\odot}$ in steps of 0.01 M$_{\odot}$. The maximum mass we
obtained for the degenerate core of an RGB star is $0.47$ M$_{\odot}$, thus we did
not consider post-flash sdB stars above this mass. The H-envelope masses,
$M_{\rm env, 0}$, considered are 0.0001, 0.0003, and 0.0006 M$_{\odot}$, where
$M_{\rm env, 0}$ is defined as the total mass of the hydrogen content directly
after the removal of the envelope. Thus, we have 18 sdB evolution tracks, which
we followed until the end of He-core burning.  After each $10^7$ yr of sdB
evolution, the seismic properties were calculated, and we only considered models
with unstable modes. We have not found unstable modes in our post-He-core
burning models. Therefore our analysis is limited to sdB stars in their He-core burning
phase, leading to a total of 402 seismic models in the range of $T_{\rm
eff}=25,000-34,000$ K and $\log g=5.4-6.0$.

The grid of non-canonical sdB stars (hereafter called grid $B$) consists of 5
tracks: ($M_{\rm sdB}$ (M$_{\odot}$), $M_{\rm env, 0}$ (M$_{\odot}$)) = (0.44,
0.005), (0.45, 0.005), (0.46, 0.005 ), (0.47, 0.0075 ), and (0.47, 0.005). Along
these tracks we have in total 98 seismic models, again taken after each $10^7$
yr of sdB evolution. In Table~\ref{gridAB}, more details about the models are given.

\begin{figure*}
   \begin{center}
  \includegraphics[angle=-90, width=18cm]{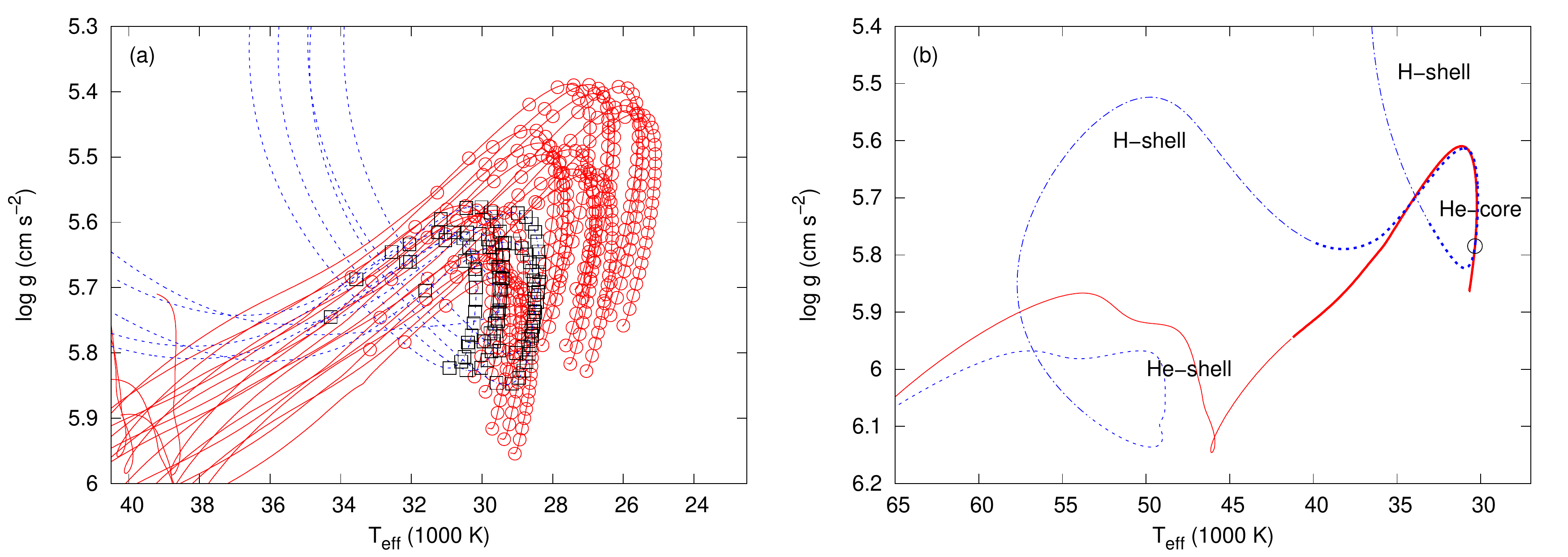}
   \caption{(a) sdB Evolutionary tracks in the $\log g-T_{\rm eff}$
   diagram. The solid curves correspond to the post-flash sdB models, and the
   dotted curves to the post-non-degenerate sdB models. The circles and squares
   indicate models on which we performed pulsation calculations for the
   post-flash and post-non-degenerate tracks respectively. (b) Evolutionary
   tracks of a post-flash sdB star created from of low-mass ($M_{\rm ZAMS}=1.00$
   M$_{\odot}$), and a post-non-degenerate sdB star created from a high mass
   ($M_{\rm ZAMS}=3.00$ M$_{\odot}$) progenitor, given by the solid and dotted
   curve respectively. Along the tracks the dominant energy source (either
   H-shell-burning, He-core-burning or He-shell-burning) is noted. The bold part
   of the tracks indicate the He-core-burning phase and the thin part the
   He-shell burning phase. During the dot-dashed parts of the
   post-non-degenerate track, H-shell burning is the dominant energy source. At
   the circle in (b), we selected from both tracks a model for a detailed
   comparison, see Section \ref{structure}.}
              \label{evtracks}
              \end{center}
    \end{figure*}

The sdB evolution tracks and the seismic models can be seen in
Fig.~\ref{evtracks}. The tracks start directly after the removal of the
envelope. For the post-flash models, this corresponds to the zero-age EHB. The
post-non-degenerate models have hydrogen extending to deeper layers, hence
allowing some H-shell-burning, before reaching the zero-age EHB. Note that, although
the two different types of sdB stars can evolve through the same $\log
g-T_{\rm eff}$ during core-He-burning, the post-He-core-burning evolution
differs, see Fig.~\ref{evtracks}b. The post-non-degenerate sdB star again has a short phase ($\sim$$10^6$
yr) of H-shell burning, before starting He-shell burning.

%______________________________________________ 
\begin{table}
\begin{minipage}{\columnwidth}
\caption{The models in grid $A$ and $B$. \protect\footnote{The label of the track given in the first column is used for reference in Fig.~\ref{chi2}. The second column gives the number of seismic models along that track. The final column shows the mass of the ZAMS models from which the sdB star is created. }}
\label{gridAB}
\begin{center}
\begin{tabular}{lllll}
\hline
\vspace{-0.25cm}\\
 track&\# models& $M_{\rm sdB}$ (M$_{\odot}$) & $M_{\rm env,0}$ (M$_{\odot}$)  & $M_{\rm ZAMS}$ (M$_{\odot}$)\\
$A1$ & 27 & 0.42 & 0.0001 &  1.85 \\
$A2$ & 27& 0.42 & 0.0003 & 1.85 \\
$A3$ & 27 & 0.42 & 0.0006 & 1.85 \\
$A4$ & 25 & 0.43& 0.0001 & 1.80 \\
$A5$ & 25 & 0.43 & 0.0003 & 1.80 \\
$A6$ & 25 & 0.43 & 0.0006 & 1.80 \\
$A7$ & 23 & 0.44& 0.0001 & 1.75 \\
$A8$ & 23 & 0.44 & 0.0003 & 1.75 \\
$A9$ & 23 & 0.44 & 0.0006 & 1.75 \\
$A10$ & 21 & 0.45& 0.0001 & 1.65 \\
$A11$ & 21 & 0.45 & 0.0003 & 1.65 \\
$A12$ & 21 & 0.45 & 0.0006 & 1.65 \\
$A13$ & 20 & 0.46& 0.0001 & 1.55 \\
$A14$ & 20 & 0.46 & 0.0003 & 1.55 \\
$A15$ & 20 & 0.46 & 0.0006 & 1.55 \\
$A16$ & 18 & 0.47& 0.0001 & 1.00 \\
$A17$ & 18 & 0.47 & 0.0003 & 1.00 \\
$A18$ & 18 & 0.47 & 0.0006 & 1.00 \\
$B1$ & 23 & 0.44 & 0.005 &  2.75 \\
$B2$ &18 &  0.45 & 0.0075 & 2.75 \\
$B3$ & 21 & 0.46 & 0.005 & 2.90 \\
$B4$ & 18 & 0.47& 0.0075 & 2.90 \\
$B5$ & 18 & 0.47 & 0.005 & 3.00 \\
\vspace{-0.25cm}\\
\hline
\vspace{-0.25cm}\\
\hline
\end{tabular}
\end{center}
\end{minipage}
\end{table}

%______________________________________________ 

%______________________________________________ 

\subsection{Comparing two representative models}\label{structure}
\begin{figure*}
   \begin{center}
  \includegraphics[angle=-90, width=18cm]{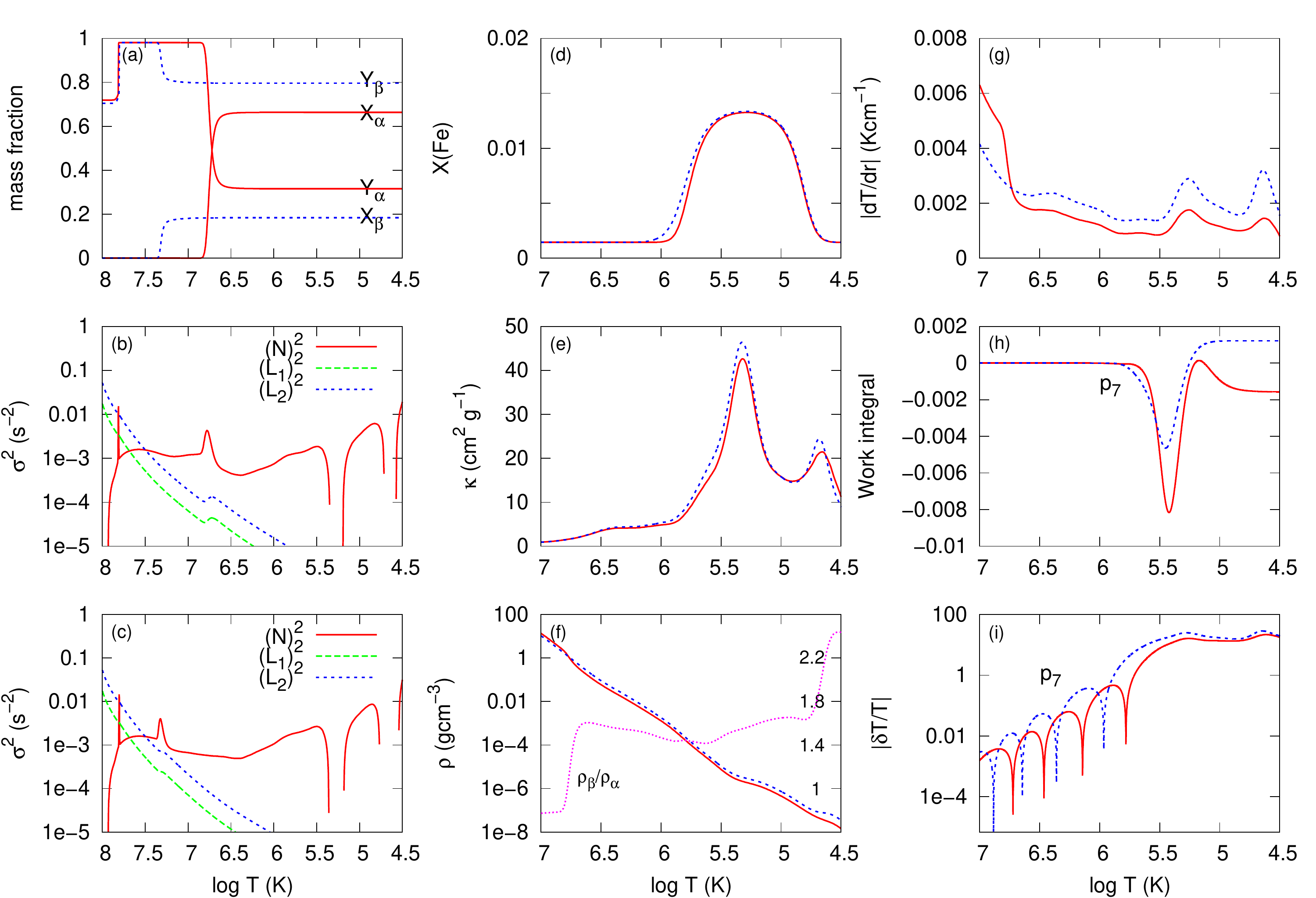}
   \caption{Physical and seismic quantities of the two representative models
   $\alpha$ (post-flash) and $\beta$ (post-non-degenerate). The profiles are shown as a function of the temperature: (a) the hydrogen ($X$) and
   helium ($Y$) mass fractions; (b)\&(c) the Brunt-V\"ais\"ala and Lamb frequencies
   for model $\alpha$ and $\beta$, respectively; (d) the iron mass fraction; (e)
   the opacity; (f) the density, we also plotted the density ratio
   $\rho_{\beta}/\rho_{\alpha}$ on the right axis; (g) the temperature gradient $|\textrm{d}T/\textrm{d}r|$; (h) the work integral for
   radial mode $p_7$; and (i) the eigenfunction $|\delta T/T|$ for the radial
   mode $p_7$.  In all panels except (b)\&(c), solid lines represent model
   $\alpha$ and dotted lines model $\beta$. Note that panels (a)-(c)
   give the profiles throughout the entire star, while the profiles in panels
   (d)-(h) are for the stellar envelope.}
              \label{detail}
              \end{center}
    \end{figure*}

We examined the physical differences in the interior structure of a post-flash
($\alpha$) and a post-non-degenerate ($\beta$) sdB with same $\log g$ and $T_{\rm
eff}$. We chose as representative the models circled in
Fig.~\ref{evtracks}b at $\log g = 5.78$ and $T_{\rm eff} = 30$ kK. One of the
main differences is the abundance profiles, see Fig.~\ref{detail}a. The He-H
transition layer of $\beta$ is much broader and located deeper in the star. The
envelope of $\beta$ is in the region where the shrinking convective core passed through during the MS, hence
the low H-abundance here: $X=0.18$. For $\alpha$, the helium core has grown into
the region that used to be part of the convective envelope during the RGB, and,
as a result, the He-H transition region is much narrower while the H-abundance
in the envelope is around $X=0.66$. We will discuss the possibility of diffusion 
of hydrogen to the surface in \S \ref{discussion}. In Fig.~\ref{detail}d, we show the iron mass
fractions, and in Fig.~\ref{detail}e the resulting opacity profiles. The outer
opacity bump near $\log T = 4.7$ is associated with helium ionization, and the iron
opacity bump near $\log T=5.3$ is enhanced by the local iron accumulation.

Two important quantities in stellar pulsation theory are the the
Brunt-V\"ais\"ala frequency $N$ and the Lamb frequencies $L_l$,
\begin{equation}
N^2 = \frac{Gm}{r^2}\frac{\delta}{H_p}\Big(\nabla_{\rm ad}-\nabla+\frac{\varphi}{\delta}\nabla_{\mu}\Big) 
\end{equation}
\[
\textrm{with}\quad\delta = -\frac{\partial\ln \rho}{\partial\ln T}\quad\textrm{and}\quad\varphi = \frac{\partial\ln\rho}{\partial\ln\mu};
\]
\begin{equation}
L_l^2 = \frac{l(l+1)c_s^2}{r^2},
\end{equation}
where $\mu$ is the molecular weight and $c_s$ is the adiabatic sound
speed. When $N^2<0$, the Ledoux criterion for dynamical stability is
violated \citep{ledoux1947}. Thus, in Fig.~\ref{detail}b\&c, the convective
regions can be clearly identified. The innermost one is related to the
convective core, and the outer two are the narrow convective layers due to iron
(near $\log T=5.3$) and helium ionization (near $\log T=4.6$). Also, chemical
gradients are apparent in $N^2$ in the form of localized peaks. The peak near
the center is identified with the C-O/He transition zone and the outermost peak
to the He-H transition zone.  The Lamb frequencies are plotted in the same
figure to indicate the propagation zones of the $g$-modes,
$\sigma^2<(N^2,L_l^2)$, and the $p$-modes, $\sigma^2>(N^2,L_l^2)$, where
$\sigma$ is the angular pulsation frequency. It is apparent that $g$-modes are
deep interior modes, while $p$-modes probe the superficial outer layers as
pointed out by \citet{charpinet2000}.

\begin{figure}
   \begin{center}
  \includegraphics[angle=-90, width=9cm]{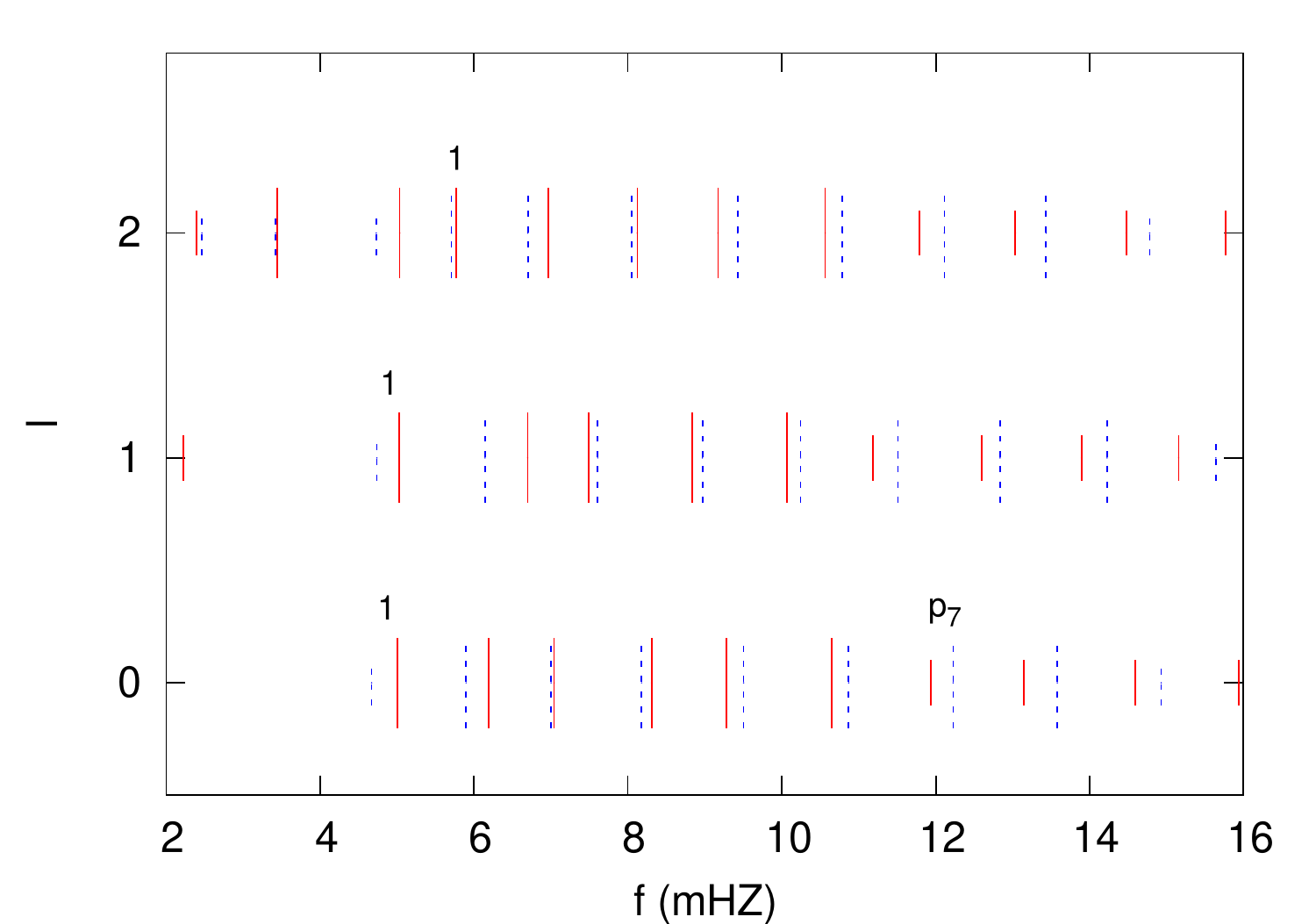}
   \caption{The pulsation frequencies for modes $l=0-2$. Solid line segments are for model $\alpha$ and dotted ones for model $\beta$. Unstable modes are given by long line segments, and stable modes by short line segments. The modes with radial order $n=1$ and radial mode $p_7$ are indicated.}
              \label{detailmode}
              \end{center}
    \end{figure}

Clearly, models $\alpha$ and $\beta$ have very different physical
characteristics. To establish how this affects their seismic properties, we
compare their frequencies in Fig.~\ref{detailmode}. Since the large frequency
separation $\Delta f = f_{n,l}-f_{n-1,l}$ is mainly dependent on the dynamical
timescale, we see that $\Delta f$ at high frequencies is more or less the same
for the two models. The lower frequencies, however, are in fact mixed modes
that are more sensitive to the core, and we see a better distinction between
models $\alpha$ and $\beta$ here.  Moreover, the frequency ranges of excited
$l=0-2$ modes are not the same for these two models; for model $\alpha$ this
range is [3.4 mHz, 10.6 mHz], and for model $\beta$, it is [5.7 mHz,14.2
mHz]. The excited modes thus have lower frequencies in model $\alpha$ than in
model $\beta$.  To understand this, we compare in Fig.~\ref{detail}h the work
integral of these two models for the radial mode $p_7$. The work integral
increases towards the surface in the driving region and decreases towards the
surface in damping regions. The surface value is the dimensionless growth rate,
positive for unstable modes and negative for stable ones. We see that $p_7$ is
unstable in model $\beta$, but stable in model $\alpha$, which can also be seen
in Fig.~\ref{detailmode}.

A first possible origin of the differences could come from the opacity, since
the driving is a $\kappa$-mechanism operating in the iron opacity bump.
Fig.~\ref{detail}e shows that the opacity is slightly larger for model
$\beta$. The driving is thus a little more efficient in model $\beta$.  But this
is not the main source of differences. Since the envelope H-fraction is much
smaller in model $\beta$ ($X=0.18$) than in model $\alpha$ ($X=0.66$), the
molecular weight is larger and, at given temperature, the density is
significantly higher ($\sim$$1.5\times$) in model $\beta$, as shown by
Fig.~\ref{detail}f. The driving of the modes is related to the opacity, which is
mainly a function of temperature. Hence, if the eigenfunctions of two given
modes have the same shape as a function of temperature, the driving is the same.
Here we compare the modes $p_7$ of two models with the same radius. As it is
usually found for $p$-modes, their last node is located at the same geometrical
distance from the surface $\Delta r$.  But the gradient of temperature is not
the same for the two models: $|\textrm{d} T/\textrm{d} r|\propto \kappa\rho$ is greater in model
$\beta$ because of the higher density, as can be seen in Fig.\ref{detail}f\&g. Hence, the difference between the
temperature at the last node and at the surface, $|\Delta
T|\,\simeq\,|\textrm{d}T/\textrm{d}r|\,\Delta r$, is greater in model $\beta$ than in model
$\alpha$. This is exactly what we find in Fig.~\ref{detail}i, where the
eigenfunction $|\delta T/T|$ is given. In terms of the temperature, the last
node is closer to the surface in model $\alpha$ than it is in model $\beta$. To
get the same driving as in model $\beta$, the last node of model $\alpha$ would
have to be deeper in the star, which is only possible by considering a mode with
lower radial order and frequency. Hence, the frequencies of excited modes are
lower in model $\alpha$ than in model $\beta$.

%______________________________________________ 

%______________________________________________ 

\subsection{Comparing two grids of models}\label{grid}
We investigated if it is possible to distinguish between a post-He-flash ($a$)
and a post-non-degenerate sdB stellar model ($b$) from observed oscillation
modes. Imagine we observed the frequencies of $b$; is it then possible to find
an acceptable seismic match in our grid of canonical post-flash models (grid
$A$)? We took as `observed' frequencies those  of unstable modes up to
$l=2$. We did this for each model $b$ in grid $B$, thus finding the best seismic
match within grids $A$ and $B$.

Since frequency separations follow from asymptotic relations for $p$-modes, the
frequency is a natural quantity for model comparison. Despite this, periods have
been used more often in the literature so far, when comparing observed modes of
sdB stars with those predicted by models. We also considered period matching,
but found frequency matching more suitable to compare the $p$-modes of the
models. This will be different for $g$-modes, where the mode period is the
natural quantity to comparing observations with models.

To quantify
`acceptable', we used the merit function
\begin{equation}
\mathcal{M}^2 = \frac{1}{n_{b}}\sum_{i=1}^{n_{b}}(f^{i}_{a}-f^{i}_{b})^2,
\end{equation}
where $f_{b}^i$ is one of the $n_{b}$ excited frequencies of star $b$, and $f_{a}^i$ is the correspondingly matched frequency of star $a$, expressed in mHz. The
frequency matching is done such that $\mathcal{M}^2$ is minimized by brute-force
fitting.  It is clear that the lower $\mathcal{M}^2$, the better the match
between $a$ and $b$. Matches with $\mathcal{M}^2>0.05$ are considered
unacceptable, which is a generous limit, as we will see later.  We investigated
four different scenarios:
\begin{itemize}
\item[(i)] We are not able to identify the modes, $\log g$ and $T_{\rm eff}$ of
the `observed' star $b$ are unknown, and the `observed' frequencies are allowed
to be matched with both stable and unstable frequencies of the `theoretical'
model $a$.
\item[(ii)] Same as (i), except the modes are identified, thus the $l$-value
must be matched.
\item[(iii)] Same as (i), except $\log g$ and $T_{\rm eff}$ are known within
errors of $d \log g=0.1$ and $d T_{\rm eff}=1000 K$.
\item[(iv)] Same as (i), except the `observed' frequencies are only matched with
unstable `theoretical' frequencies, i.e.~assuming 
that the theory correctly predicts which frequencies are exited and which are 
not.
\end{itemize}

\begin{figure*}
   \begin{center}
   \includegraphics[angle=-90, width=16cm]{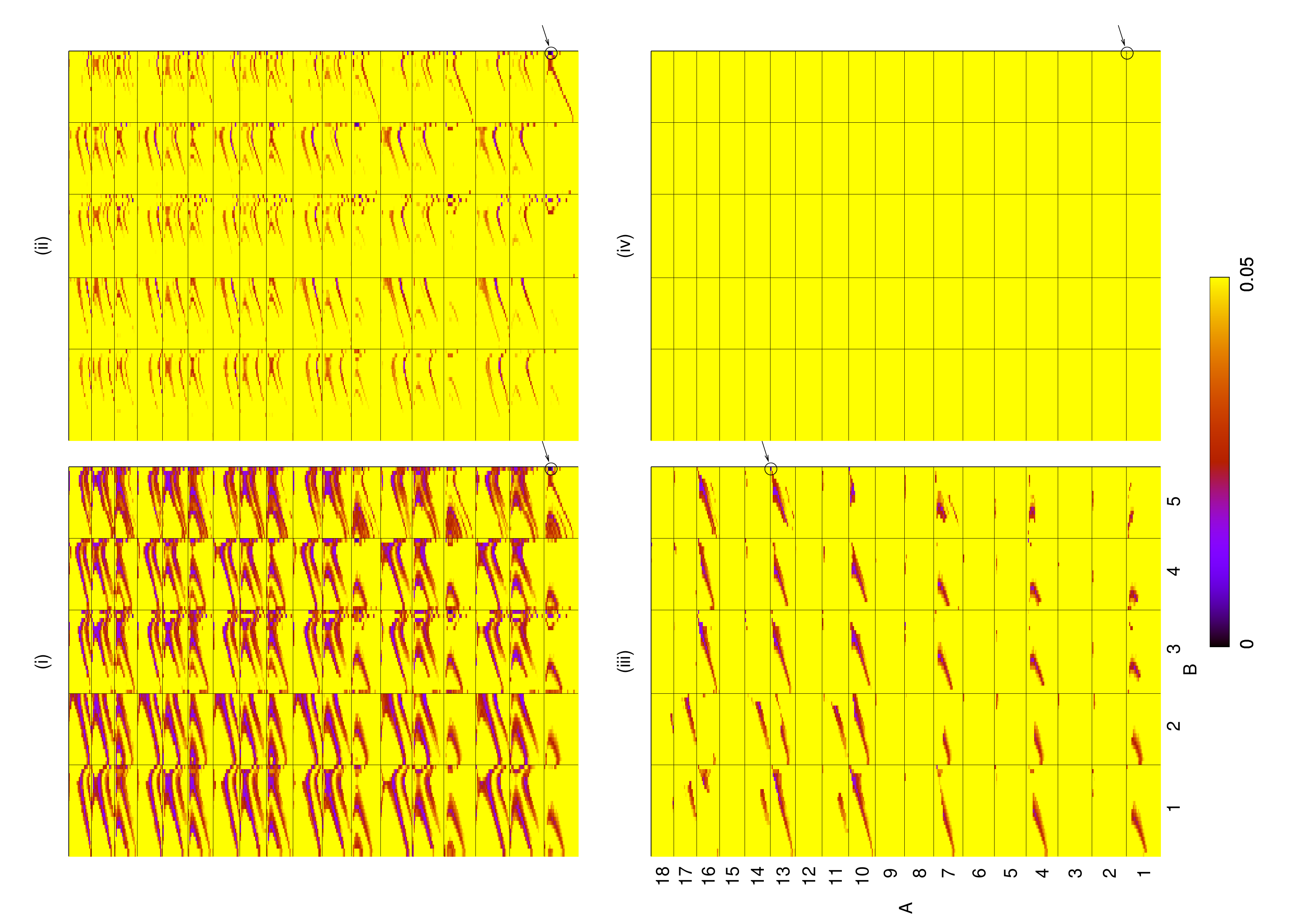}
   \caption{Colourmap of $\mathcal{M}^2$. The models of grid $A$ are plotted
   along the vertical axis, and the models of grid $B$ along the horizontal
   axis. The models are divided in blocks according to evolutionary track, where
   we have ordered the blocks with increasing mass (see Table~\ref{best}), and
   in the blocks the models are ordered with increasing age. The four panels are
   for the four different scenarios (i), (ii), (ii), and (iv), as described in the
   text. In each panel with have pointed out and circled the gridpoint with minimum
   $\mathcal{M}^2$. The frequency matching of these gridpoints can be seen in
   Fig.~\ref{lmode}. Note that panel (iv) has no acceptable matches, but we
   still circled the one with the lowest $\mathcal{M}^2$. }
              \label{chi2}
              \end{center}
    \end{figure*}

In Fig.~\ref{chi2}, we show $\mathcal{M}^2$ for each gridpoint in grids $A$ and
$B$ for the scenarios (i)-(iv).  The matches with low $\mathcal{M}^2$ are
visible as dark diagonal regions, This is a result of the change in
frequencies during the sdB evolution.  From Fig.~\ref{chi2}(i)-(iii), it is clear that the distinction between models $a$ and $b$ is drastically increased if we have either mode identification or spectroscopic $\log g$ and $T_{\rm eff}$ values. Fig.~\ref{chi2}(iv) shows that, if we only allow matching to unstable (and not stable) frequencies of $a$, there are no matches. 

\begin{figure}
  \begin{center}
  \includegraphics[angle=-90, width=8cm]{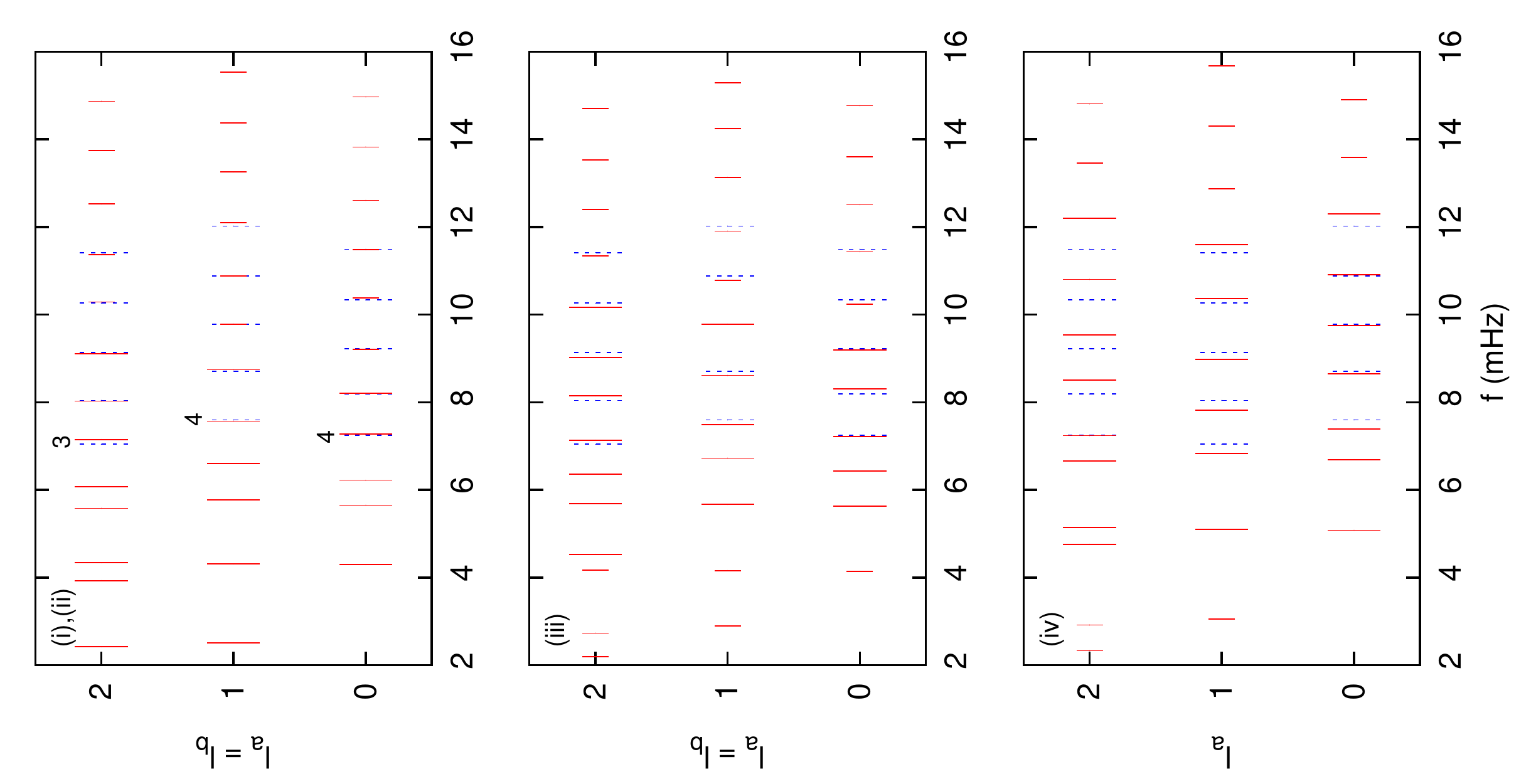}
  \caption{The frequency matches of the minimum $\mathcal{M}^2$ gridpoints.  The
$l$-value of model $a$ is depicted on the vertical axis. For scenarios (i) and (ii), this is equal to the $l$-value of model $b$, for scenario (iv) it is not. Solid line segments are for model $a$ and dotted ones for model $b$.  In the upper panel, we have indicated the lowest radial order of the unstable modes of model $b$. 
}
 \label{lmode}
  \end{center}
  \end{figure}

The matches with lowest $\mathcal{M}^2$ are circled in Fig.~\ref{chi2} and
details of these models are shown in Fig.~\ref{lmode} and Table~\ref{best}.  For
all scenarios the same model $b$ gives the best match, namely the last model of evolutionary track $B5$. We understand that in terms of only the higher frequencies with radial order $n\geq3$ being excited in this model. As we discussed in \S \ref{structure}, the lower frequencies are more sensitive to the deeper layers, thus the distinction between models $a$ and $b$ is better detected at low frequencies.

As a comparison to a real case, we considered the optimal model for \object{PG\,0014$+$067}, for which \citet{brassard2001} found $\chi^2=0.5374$, where $\chi^2$ is a merit
function based on mode period comparison. Translated to our frequency merit
function, this is equivalent to $\mathcal{M}^2 = 0.0084$. Although we find, in principle, seismic matches between $a$ and $b$ with $\mathcal{M}^2$ of this order for scenarios (i)-(iii), they are not statistically favoured. For scenario (i) we find that 12 of the 98 models in grid $B$ can be matched with a model in grid $A$ with $\mathcal{M}^2\leq0.01$, and this is 7 for scenario (ii), and only 1 for scenario (iii). 

%______________________________________________ 
%______________________________________________ 

%______________________________________________ 
%______________________________________________ 
\begin{table}
\caption{The models with minimum $\mathcal{M}^2$ for the four different scenarios.}
\label{best}
\begin{center}
\begin{tabular}{llll}
\hline
\vspace{-0.25cm}\\
&(i) \& (ii) & (iii) & (iv)\\
\vspace{-0.25cm}\\
\hline
\vspace{-0.25cm}\\
$\mathcal{M}^2$ & 0.0018 & 0.0082 & 0.080  \\
\vspace{-0.25cm}\\
\hline
\vspace{-0.25cm}\\
track &$A1$& $A13$ & $A1$ \\
$M_{\rm sdB}$ (M$_{\odot}$) & 0.42 & 0.46 & 0.42\\
$M_{\rm env, 0}$  (M$_{\odot}$) & 0.0001 & 0.0001& 0.0001\\
$\log g$ &5.69 &5.69 &5.78 \\
$T_{\rm eff} (K)$ & 29216& 33117&32187 \\
EHB age (yr) & $2.2\times 10^8$ & $1.9\times 10^8$&$2.7\times 10^8$\\
\vspace{-0.25cm}\\
\hline
\vspace{-0.25cm}\\
track & $B5$ & $B5$ & $B5$\\
$M_{\rm sdB}$ (M$_{\odot}$) &0.47  &  0.47 & 0.47\\
$M_{\rm env, 0}$ (M$_{\odot}$)  & 0.005 & 0.005 & 0.005\\
$\log g$ &5.69 & 5.69 & 5.69\\
$T_{\rm eff} (K)$ & 33558& 33558 &33558 \\
EHB age (yr) & $1.8\times 10^8$ & $1.8\times 10^8$ &$1.8\times 10^8$\\
\vspace{-0.25cm}\\
\hline
\end{tabular}
\end{center}
\end{table}
%______________________________________________ 
 
\section{Discussion \& conclusions}\label{discussion}

We studied the so far neglected, post-non-degenerate sdB stars and compared
their physical and seismic characteristics with those of canonical post-flash
sdB stars, both formed in the CEE channel. The results presented here are a
first step in distinguishing these two kinds of sdB stars on the basis of their
observed oscillation character, which is necessary if seismic modelling is to
achieve reliable mass determination. Furthermore, the observation of a
post-non-degenerate sdB star in a post-CE binary would give strong
constraints on the CE evolution. We plan to continue such investigations with an
application to the sdB pulsator in the post-CE, eclipsing binary \object{PG\,1336$-$018}
which we started in \citet{hu2007} and \citet{vuckovic2007}.

We find that, in principle, a post-non-degenerate sdB star may appear as an EC
14026 star with similar pulsation frequencies as the canonical post-He-flash sdB
star, although it is not likely. Additional observables, such as spectroscopic $\log g$ and
$T_{\rm eff}$ determinations and/or empirical mode identification from
observables enable us to distinguish the two types of sdB stars more decisively. The frequency range of the unstable modes is also an important
discriminator between the two formation channels. In general, for the same $\log
g$ and $T_{\rm eff}$ values, the excited frequencies of the post-non-degenerate
sdB star are higher than the excited frequencies of the post-flash star. This is
a direct result of the differing interior structures. Thus, special attention must be paid when observed frequencies are matched with theoretically predicted ones of modes that are not excited.

Up to now, there have not been any evolutionary models of sdB stars available that include
the coupling between diffusion and evolution consistently. This is a deficiency,
since iron accumulation due to radiative levitation is responsible for the pulsational instability in these stars \citep{charpinet1996}. Also, it has been shown by
\citet{fontaine2006} that the iron accumulation changes the frequencies
significantly. In our study, we have parametrized the iron accumulation, so that we can, at least in an approximative manner, simultaneously take into account the effects of iron enhancement and evolution on the pulsation modes. 

Here we have not considered the influence of the other diffusive processes, i.e.~diffusion due to gradients of pressure, temperature, and concentration. To a certain extent this can affect our results, because one of the main differences between the two types of sdB stars is the chemical composition of the stellar envelope. Specifically, we find in the envelope of the post-non-degenerate sdB star an H-mass fraction of $X=0.18$, while the post-flash sdB star has $X=0.66$ there. Normally, it is assumed that sdB stars have H-rich or even pure H-envelopes, caused by gravitational settling. While this is true for the outermost layers, diffusion is not expected to work efficiently at depths $\log q\gtrsim-3$ \citep{richard2002, michaud2007}. Since the envelopes of the post-non-degenerate sdB stars extend to $\log q\gtrsim -2$  (i.e.~$T\gtrsim10^7$ K), we do not expect diffusion to wash away all the qualitative differences in the chemical profiles, although the differences may be less pronounced.  
Diffusion, however, will significantly change the surface abundances of our models and likely will bring them into agreement with the observed values. Spectroscopic line profile analysis has shown that the majority of sdB is He-deficient, and only a few are He-rich \citep{edelmann2003,lisker2004}. Stellar evolution models 
that include diffusion coupled to reliable atmosphere models are needed to assess 
whether the two different formation channels will be distinguishable via a 
spectroscopic abundance analysis. We are currently computing such evolutionary sdB models including diffusion due to gradients of pressure, temperature, and concentration. Our preliminary results indeed agree with our expectations, i.e.~the H-surface abundance increases on a very short timescale, but the chemical profiles at deeper layers are not affected. The pulsational properties of these improved models will be discussed in detail in a forthcoming paper.

We have made a modest grid of models that is sufficient for our comparative
study. Detailed seismic modelling of an observed star, however, will require a
finer grid. For now, we have chosen not to make sdB models above 0.47
M$_{\odot}$, since this is the maximum mass the degenerate He-core of a red
giant with $Z=0.02$ can have before experiencing the He-flash. A metallicity of
$Z=0.004$ allows the He-core to grow up to 0.48 M$_{\odot}$ on the RGB.
However, we find that, in order to excite modes in these low
metallicity stars, an iron enhancement greater than a factor 10 is required.
This was to be expected, since \citet{charpinet1996} found unstable pulsation
modes for models with uniform $Z\geq0.04$ in the H-rich envelope. We have
therefore not pursued these models further. The question whether post-flash sdB
stars can have masses $>0.47$ M$_{\odot}$ is also closely related to the input
physics (e.g.~convective overshooting) and the physics of the He-flash, and needs to be examined further.

In this paper, we have focused on the short-period $p$-mode sdB pulsators. The
case of the long-period $g$-mode sdB pulsators is, although challenging from an
observational point of view, an additional very interesting theoretical case
study.  The $p$-modes only probe the outermost layers, and hence are less
affected by the differing composition gradients than the $g$-modes, as they
propagate deeper into the star. The long-period sdB pulsators are interpreted as
cooler sdB models with much thicker hydrogen-envelopes than the short-period sdB
pulsators \citep{fontaine2003, jeffery2006}. Since the $g$-modes are deep
interior modes, full evolutionary models including iron accumulation, as developed here, are required to model these stars. At present, these are not
available yet. We are currently developing a similar approach to the one presented here
to study the long-period sdB pulsators.

\begin{acknowledgements}
We are grateful to P.~P.~Eggleton for the use of his evolution code, and to
E.~Glebbeek and S.~de Mink for their help with this code. We thank W.~van Ham
for his help with the frequency-matching algorithm. We would also like to thank
M.~Vu\v{c}kovi\'{c}, R.~\O stensen, and M.~D.~Reed for stimulating
discussions. We also thank the referee, Z.~Han, for useful comments. HH thanks the department of Astrophysics at the University of Li\`ege
for its hospitality. HH acknowledges a PhD scholarship through the ``Convenant
Katholieke Universiteit Leuven, Belgium -- Radboud Universiteit Nijmegen, the
Netherlands''. CA acknowledges financial support from the ``Stichting Nijmeegs
UniversiteitsFonds (SNUF)'' and the Netherlands Research School for Astronomy
(NOVA). HH and CA are supported by the Research Council of Leuven University,
through grant GOA/2003/04. GN is supported by NWO-VENI grant 639.041.405.
This work was supported by the European Helio- and Asteroseismology
Network (HELAS), a major international collaboration funded by the
European Commission's Sixth Framework Programme.
\end{acknowledgements}

\bibliographystyle{aa}
\bibliography{10233}
\end{document}